\documentclass[eqsecnum,showpacs,aps,epsf,twocolumn,nofootinbib]{revtex4}

\usepackage{graphicx}

\begin{document}

\draft


\title{Creation of a brane world with Gauss-Bonnet term}

\author{
Koh-suke Aoyanagi\footnote{E-mail
address : aoyanagi@gravity.phys.waseda.ac.jp}}
\author{
Kei-ichi Maeda\footnote{E-mail address : maeda@gravity.phys.waseda.ac.jp}
}

\address{$^{1}$Department of Physics, Waseda University, 3-4-1 Okubo,
Shinjuku-ku, Tokyo 169-8555, Japan~}
\address{$^2$ Advanced Research Institute for Science and Engineering,
Waseda  University, Shinjuku, Tokyo 169-8555, Japan~}
\address{$^3$ Waseda Institute for Astrophysics, Waseda University,
Shinjuku,  Tokyo 169-8555, Japan~}

\date{\today}

\begin{abstract}
We study a creation of a brane world using an instanton solution.
We analyze a brane model with a Gauss-Bonnet term in a bulk spacetime.
The curvature of 3-brane is assumed to be closed, flat, or open.
We construct instanton solutions with branes for those models,
and calculate the value of the actions to discuss an initial state
of a brane universe. 
\end{abstract}


\maketitle

\section{INTRODUCTION}
Although a big bang universe is 
very successful, it predicts the existence of an initial  singularity.
To resolve such a difficulty, we have to search for
new gravitation theory such as quantum gravity.
However, we could so far  not  find it.
As a first step to reach it, we may consider a mini-super space
and quantize the isotropic and homogeneous universe,
which is the so-called quantum cosmology \cite{Vilenkin82,Hartle-Hawking83}.
Vilenkin claimed that the universe is created from nothing \cite{Vilenkin82}.
This approach is based on the picture
that the universe is spontaneously nucleated
in a de Sitter space.
The mathematical description of this nucleation
is analogous to a quantum tunneling 
through a potential barrier \cite{Coleman-deLuccia80}.
Another approach to quantum cosmology is developed by
Hartle and Hawking \cite{Hartle-Hawking83}, who  proposed that the
wave function of the Universe is given by a path integral 
over non-singular compact Euclidean geometries, which is called a ``no
boundary'' boundary condition. 

When we discuss the early stage of the universe, however, a unified theory of
fundamental interactions and particles will play a very important role.
Among such unified theories, a string/M theory is the most promising candidate,
which is constructed in higher-dimensional spacetime. 
Based on such higher-dimensional theories,
new cosmological scenario has been proposed, that is 
a brane cosmology.
The prototype of a brane world was first discussed in \cite{Akama,
Rubakov-Shaposhnikov}.
Combined it with the idea of D-brane found by Polchinski in a string
theory \cite{Polchinski}, a new paradigm of a brane world has been
developed \cite{Arkani}.
Randall and
Sundrum also proposed  new brane
models \cite{Randall-Sundrum99_1,Randall-Sundrum99_2}. 
In their second model \cite{Randall-Sundrum99_2}, 
it was shown that the four-dimensional gravity is recovered even in
an infinite bulk spacetime.
It also provides new terms in the Friedmann equation of 
a brane universe \cite{Binetruy-Deffayet-Langlois00}.
The effective 4-dimensional (4D)
 Einstein equations with these new terms are
obtained covariantly 
 by projection of the 5-dimensional (5D) spacetime onto the brane world
\cite{Shiromizu-Maeda-Sasaki00}, by which we understand naturally the origin of
new terms. 

If we believe such higher-dimensional cosmological scenario, 
we have to invoke how such a universe is created.
In particular, because a brane structure is highly inhomogeneous in a
higher-dimensional bulk spacetime,
we may wonder how such a brane universe is born and starts to evolve.
As for creation of a
brane universe, a few works has been so far done.
Garriga and Sasaki first constructed 
an inflating brane instanton
of Randall-Sundrum model \cite{Garriga-Sasaki00}.
This instanton is obtained by gluing two spherical parts of $\mbox{AdS}_5$.
Hawking, Hetrog and Reall
consider the creation using an instanton,
and discuss  inflation and 
a fluctuations during the de Sitter phase in the model which
contains the quantum correction term called a trace anomaly on the brane
\cite{Hawking-Hetrog-Reall01,Hawking-Hetrog-Reall00}.
The trace anomaly term is first discussed by Starobinsky
 for inflation \cite{Starobinsky80}.
Another quantum correction term called induced gravity
is introduced on the brane
 by Dvali, Gabadadze, and Porrati
\cite{Dvali-Gabadadze-Porrati00}.
For this model, the effective gravitational equations on the brane are
derived \cite{Maeda-Mizuno-Torii03}.
This may provide new de Sitter phase without a cosmological constant.

When we discuss such quantum effects on the brane,
we may also have to include quantum effects in the bulk.
In the higher
dimensional theory,  the higher curvature
correction terms should be added to the Einstein-Hilbert action.
These terms appear in  the low energy effective action 
of string theory via quantum one-loop corrections.
In fact the low energy effective actions of some sting theories
include ${\mathcal R}^{ABCD}{\mathcal R}_{ABCD}$
interactions, but this term gives rise to a ghost.
In order to resolve this problem, 
Zwiebach \cite{Zwiebach85} (also see \cite{Zumino86}) introduce a 
ghost-free Gauss-Bonnet combination.
Hence, the Gauss-bonnet term should be included when we 
discuss a brane universe with some quantum corrections.

For RS II type model in the Einstein-Gauss-Bonnet theory,
it is shown that the graviton zero mode is localized
at low energies 
as in the original RS II model and that the correction of the Newton's law 
becomes milder by including the Gauss-Bonnet term 
\cite{Deruelle-Sasaki03,Kim-Lee01}.
Also the covariant effective 4D Einstein equation is given by
Maeda and Torii \cite{Maeda-Torii03}.

In this paper we consider the creation of a brane universe 
using an instanton solution 
in the Einstein-Gauss-Bonnet theory.
We construct a brane instanton 
including a Gauss-Bonnet term.
In \S \ref{action-eqm}, we present Euclidean action 
and Euclidean equations of motion
in the Einstein-Gauss-Bonnet theory.
In \S \ref{instanton}, we obtain the instanton solutions
which 3-curvature is positive, zero or negative.
We show that only for a closed model with positive curvature there exist
instanton solutions either with a single brane or with two branes. For
a flat or an open model, there does not exist any instanton solution with
a single brane because of a boundary condition. We also calculate the
value of the Euclidean actions to discuss an initial state of a brane
universe.
In Section \ref{trace-anomaly} 
we consider the model including a trace anomaly term on the brane.
We show
there exists instanton solution with an inflating brane.
In \S \ref{conclusion}, conclusion follows.

\section{ACTION AND EQUATIONS OF MOTION} \label{action-eqm}
The Euclidean action for the brane world with Gauss-Bonnet
term in the 5D space time is described by two parts: One in a bulk
spacetime (${\cal M}$)  and the other in brane boundary
hypersurfaces ($\partial{\cal M}=\sum_i\partial{\cal M}_i$), i.e.
\begin{eqnarray}
 S_E=S_E^{\rm bulk}+S_E^{\rm brane}\,.
\end{eqnarray} 
The bulk action is given by 
\begin{eqnarray}
 S_E^{\rm bulk}=- \frac{1}{2\kappa^2_5} \int_{\mathcal M} dx^5 \sqrt{g}
 \left[{\mathcal R}  -2 \Lambda + \alpha {\mathcal L}_{GB} \right],
\end{eqnarray}
where
\begin{eqnarray}
 {\mathcal L}_{GB}= {\mathcal R}^2 -4 {\mathcal R}_{AB}{\mathcal R}^{AB}
  + {\mathcal R}_{ABCD}{\mathcal R}^{ABCD} .
\end{eqnarray}
$\kappa_5^2$ is the 5D gravitational constant, $\Lambda$
is a cosmological constant,  ${\mathcal R}, {\mathcal R}_{AB}$
and ${\mathcal R}_{ABCD}$ are the 5D scalar curvature, Ricci tensor
and Riemann tensor, respectively, and $\alpha$ is a constant,
which is related to a string coupling constant. The induced
4D metric $h_{\mu \nu}$ on a 3-brane is defined by 
\begin{eqnarray}
 h_{AB} = g_{AB} - n_A n_B,
\end{eqnarray}
where $n_A$ is the spacelike unit vector field which normal to the brane
hypersurface. The action of the branes is given by the following form:
\begin{eqnarray}
 S_E^{\rm brane} = -\sum_{i}  \int_{\partial
{\mathcal M}_i} d^4x
 \sqrt{h}  \left[ \frac{1}{\kappa_5^2} L_{\rm surface}(\partial {\mathcal M}_i)-
\lambda_i
\right],\,\,\,\,
\end{eqnarray}
where 
\begin{eqnarray}
 L_{\rm surface}=K +2\alpha (J-2G^{\rho \sigma}K_{\rho \sigma})
\end{eqnarray}
is a surface term of the 5D gravitational action
\cite{Myers87,Gibbons-Hawking77,Chamblin-Reall99}.
$\lambda_i$ is a tension on the $i$-th brane,
$K_{\mu \nu}$ is the extrinsic curvature of a brane, 
$K=K^{\mu}_{\mu}$, and $G_{\mu \nu}$ is the Einstein tensor of the
induced metric $h_{\mu \nu}$.
$J$ is a trace of $J_{\mu\nu}$, which is given by some combination of
the extrinsic curvature defined by
\begin{eqnarray}
 J_{\mu \nu}&=&\frac{1}{3} \left(2KK_{\mu \rho}K^{\rho}_{\ \nu}
 + K_{\rho \sigma}K^{\rho \sigma}K_{\mu \nu} \right.
\nonumber \\
&&\left.-2K_{\mu \rho}K^{\rho
 \sigma}K_{\sigma \nu} -K^2 K_{\mu \nu}   \right).
\end{eqnarray}

The total action $(S_E=S_E^{\rm bulk}+S_E^{\rm brane})$ gives the field 
equations as
\begin{eqnarray}
  {\mathcal G}_{AB}+\alpha {\mathcal H}_{AB}= -\Lambda g_{AB} -\sum_i\lambda_i
   g_{AB} \delta(\partial {\mathcal M}_i),
\end{eqnarray}
where
\begin{eqnarray}
 {\mathcal G}_{AB}={\mathcal R}_{AB}-\frac{1}{2}g_{AB}{\mathcal R},
\end{eqnarray}
and
\begin{eqnarray}
 {\mathcal H}_{AB}&=&2\left[{\mathcal R}{\mathcal R}_{AB}
 -2{\mathcal R}_{AC}{\mathcal R}^{C}_{\ B}-2{\mathcal
R}^{CD}{\mathcal R}_{ACBD} \right. \nonumber \\
&& \left.+{\mathcal R}_A^{\ CDE}{\mathcal R}_{BCDE} 
\right] 
-\frac{1}{2}g_{AB}{\mathcal L}_{GB}.
\end{eqnarray}

Since we are looking for an instanton solution, we assume a highly
symmetric Euclidean spacetime, which metric is given by
\begin{eqnarray}
 ds_E^2 = dr^2+ b(r)^2 \gamma_{\mu \nu}dx^{\mu}dx^{\nu}, \label{Emet}
\end{eqnarray}
where $\gamma_{\mu \nu}$ is a 4D metric with maximal
symmetry. This maximally symmetric 4D space is classified into three
cases by the signature of curvature, i.e. $k=$0 (zero), 1 (positive), or
$-1$ (negative). It corresponds to the curvature sign of the
Friedmann universe after creation.
Since the Euclidian space must be compact when we discuss its creation,
in the case of $k$=0 or $-1$, we have to make a space compact by identification.
Then the flat  spacetime is a  4D torus, and that with $k=-1$ has more
complicated topology.
Although the spacetimes are compact,  we shall call them ``flat'' and
``open'' for $k=0$ and $-1$   as well as ``closed'' for $k=1$.

The equations of motion under the above ansatz are given by
\begin{eqnarray}
&& 3\left({b''\over b}-{k-b'^2\over b^2} \right)
 + 12\alpha \frac{(k-b'^2)}{b^3} b'' 
\nonumber \\
&& ~~~~~~~~~~~~~~= -\Lambda -\sum_i\bar{\lambda}_i \delta(r-r_i), \label{eqm1}
\end{eqnarray}
\begin{eqnarray}
 6\frac{(k-b'^2)}{b^2}\left\{1+ 2\alpha \frac{(k-b'^2)}{b^2} \right\}=
\Lambda, 
\label{eqm2}
\end{eqnarray}
where the prime denotes the derivative with respect to $r$ and 
$\bar{\lambda}_i=\kappa_5^2{\lambda}_i$.

\begin{figure}[t]
\begin{center}
\begin{tabular}{c}
 \includegraphics[height=4cm]{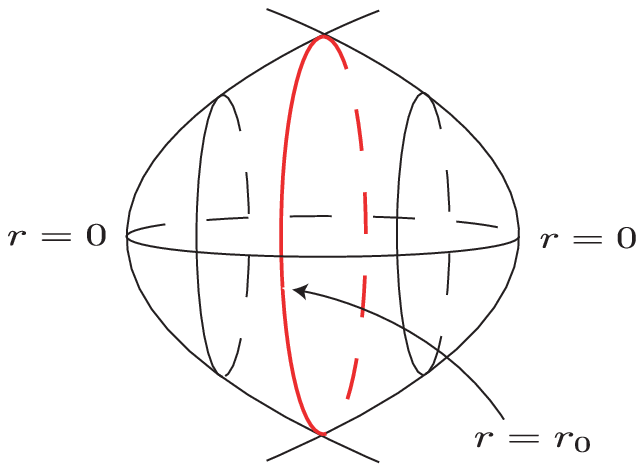} \\
 (a) \\[5mm]
 \includegraphics[height=4cm]{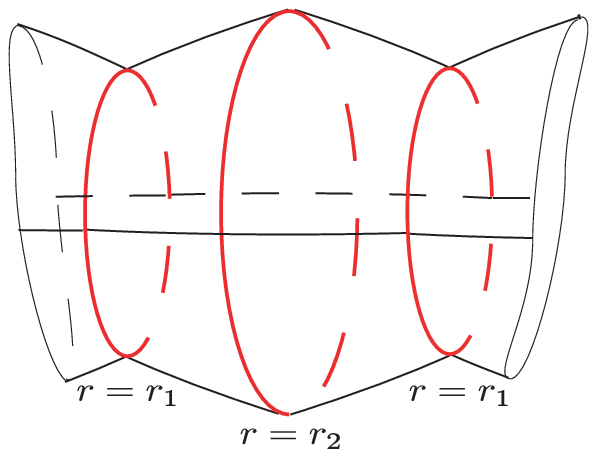} \\
 (b) \\
\end{tabular}
 \caption[fig1]{brane instanton. The thick vertical
circle at $r=r_i$ represents the 4-sphere brane at which
the two identical 5-dimensional anti-de Sitter spaces are glued.
} \label{fig1}
\end{center}
\end{figure}

By integrating the first equation for a small interval 
$(r_i-\epsilon, r_i +\epsilon)$
including a brane,  one obtain 
the junction condition \cite{Israel66} at $r=r_i$ as
\begin{eqnarray}
 \left[ \frac{b'}{b}\left\{3-4 \alpha \left({b'^2\over b^2}-{3 k\over b^2}\right) 
 \right\}\right]^{\pm} = - \bar{\lambda}_i \,,
\end{eqnarray}
where $[ \ ]^{\pm}$ denotes
\begin{eqnarray}
 [ A ]^{\pm}=A(r_i+\epsilon) -A(r_i -\epsilon )
 \equiv A_{+}-A_{-}
\end{eqnarray}
With the ansatz of $Z_2$ symmetry, which gives the relation of  
$[A]^{\pm}=2A_{+}=-2A_{-}$, we obtain
\begin{eqnarray}
 \frac{b'}{b}\left\{3-4 \alpha \left({b'^2\over b^2}-{3 k \over b^2}
\right) \right\}= \mp \frac{\bar{\lambda}_i}{2} , 
 \label{junc}
\end{eqnarray}
Here the upper (lower) sign is applied at $r=r_2$ (at $r=r_1$)
for a two-brane model.
For a single brane model, we apply the upper case at $r=r_0$.
Through this paper, we use the notation $r_i (i=0,1,2)$ where
$i$ means the number of a brane. 
For a two-brane model $i=1$ and $2$ stand for
a negative and a positive  tension brane, respectively.
While for a single brane model, 
we use $i=0$ standing for a brane (see Fig. \ref{fig1}).

With equations of motion the action is reduced to
\begin{eqnarray}
&& S_E=- \frac{8V_4^{(k)}}{\kappa^2_5} \int dr 
 \left\{ \Lambda+ 2\frac{b''}{b} -3\frac{k-b'^2}{b^2} \right\}
 \nonumber \\
&&
+\frac{V_4^{(k)}}{\kappa_5^2} \sum_{i} 
 \left[\frac{b'}{b}\left\{1-12\alpha \left(\frac{b'^2}{b^2} -
	 \frac{3k}{b^2}\right)\right\}\right]^{\pm}_{r=r_i},
\end{eqnarray}
where $V_4^{(k)}$ is a volume of 4D manifold
with the curvature $k$. 

\section{INSTANTON SOLUTION} \label{instanton}

We first provide a solution of Eqs. 
(\ref{eqm1}) and (\ref{eqm2}) in a bulk.
Eq. (\ref{eqm2}) gives the quadratic equation,
\begin{eqnarray}
 6X+12\alpha X^2 = \Lambda , \label{qe1}
\end{eqnarray}
where
\begin{eqnarray}
 X \equiv \frac{k-b'^2}{b^2}. \label{eq1}
\end{eqnarray}
If $\alpha \neq 0$, Eq. (\ref{qe1}) gives two solutions;
\begin{eqnarray}
 X = X_{\pm}\equiv \frac{-1 \pm \eta}{4\alpha},
\label{sol1}
\end{eqnarray}
where $\eta=\sqrt{1+4\alpha \Lambda/3}$.
Here we find two solutions; one is called as a plus-branch and the other is a
minus-branch.
The limit of $\alpha\rightarrow o$ exists only for the plus-branch solution.

In this paper we consider only a negative cosmological constant 
($\Lambda < 0$).
\footnote{In the case of $\Lambda=0$, there also exist instanton solutions. For
$k=1$, the minus-branch solution is included in (\ref{sol1}), and the plus-branch
solution is also included in the limit of $X_+ \rightarrow 0$. This plus-branch
solution is 5D Milne universe. For $k=0$ the solutions of both branches is
included in (\ref{sol2}). For $k=-1$ only the minus-branch solution exists and is
included in (\ref{sol3}).}
Then we have the following constraint,
\begin{eqnarray}
  -\frac{3}{4\alpha} \leq \Lambda <  0 \,,
 \label{cond1}
\end{eqnarray}
which is required in order that $X_{\pm}$ is
a real value. The range of $\eta$ is restricted as $0 \leq \eta <1$ 
from the constraint (\ref{cond1}).
Under this condition, $X_{\pm}$ is always
negative. We then introduce a typical length scale as
\begin{eqnarray}
l_\pm\equiv (-X_\pm)^{-1/2}=
\left[-{3(1\pm \eta)\over \Lambda}\right]^{1/2}\,.
\end{eqnarray}
Since $X'=0$, we find $b''=l_\pm^2b$ from Eq. (\ref{eq1}). This with Eq.
(\ref{qe1}) guarantee Eq. (\ref{eqm1}).
Hence Eq. (\ref{sol1}) gives a bulk solution.

In what follows, we discuss the instanton solutions 
for each value of $k$ in order.


\subsection{de Sitter brane instanton $(k=1)$} \label{deSitter}
In the case of $k=1$, the solution (\ref{sol1}) is
written by,
\begin{eqnarray}
 b(r) = l_\pm \sinh\left({r\over l_\pm}\right)\,,
\end{eqnarray} 
which  also satisfies Eq. (\ref{eqm1}) in a bulk.
From this bulk solution, we construct an instanton solution
by cutting the space at $r= r_i$ and
gluing two copies of it on the surfaces of the
corresponding point in order that 
a compact Euclidean manifold (instanton) is obtained.
At $r=r_i$, we impose the Israel's 
junction condition (\ref{junc}) with $k=1$.
For a single brane instanton, we impose 
``no boundary boundary condition'' at the origin \cite{Hartle-Hawking83}. 
For a two-brane model, we impose the junction condition
at $r_1$ and $r_2$.
As a result, the tension of $i$-th brane is determined by
these junction conditions.
Substituting (\ref{sol1}) into (\ref{junc}),
\begin{eqnarray}
 \bar{\lambda}_i^{(\pm)}& = &(-1)^i \frac{2}{l_\pm} 
 \left[(2\pm \eta) \frac{\cosh(r_i/l_\pm)}{\sinh(r_i/l_\pm)}
\right.
 \nonumber \\ &&
\left. ~~~~~~~~~~+2(1 \mp \eta)
\left(\frac{\cosh(r_i/l_\pm)}{\sinh^3(r_i/l_\pm)}
 \right) \right]\,.
\end{eqnarray}
When we take the limit of  $\alpha \rightarrow 0$ in the plus-branch,
we recover the Garriga-Sasaki instanton, i.e.
$l_\pm=l\equiv\sqrt{-6/\Lambda}$, $\eta=1$ , and 
\begin{eqnarray}
 \bar{\lambda}_i= (-1)^i \frac{6}{l} \coth \left(\frac{r_i}{l}
\right) \,.
\label{tens1}
\end{eqnarray}

Here we note a `critical' tension.
In a brane model with the Gauss-Bonnet term in a bulk, we
find some contribution from the Gauss-Bonnet term in
a 4-dimensional cosmological constant.
As a result, the fine-tuned value of the tension 
to find 
the 4D Minkowski brane, which we shall call 
a critical tension,
 is modified from the
Randall-Sundrum's value. 
The fine-tuned value is
given by \cite{Maeda-Torii03}
\begin{eqnarray}
 \alpha \bar{\lambda}^{2}_{\rm cr}=1-4 \alpha \Lambda \mp 
 \left(1+\frac{4}{3}\alpha \Lambda \right)^3 
\label{cr_ten}\, .
\end{eqnarray}

\begin{figure}[ht]
\begin{center}
\begin{tabular}{c}
 \includegraphics[height=5.5cm]{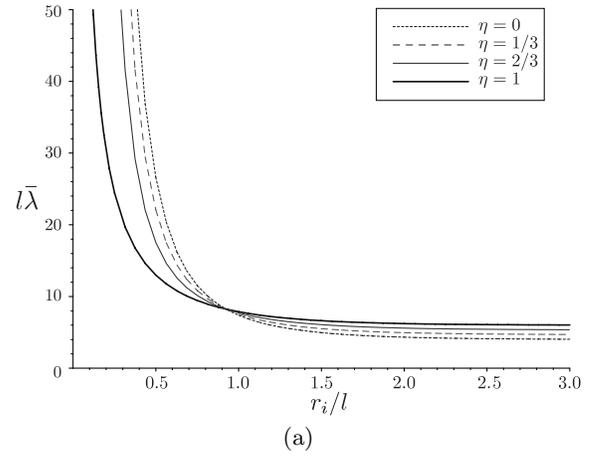} \\ 
 (a)\\[5mm]
 \includegraphics[height=5.5cm]{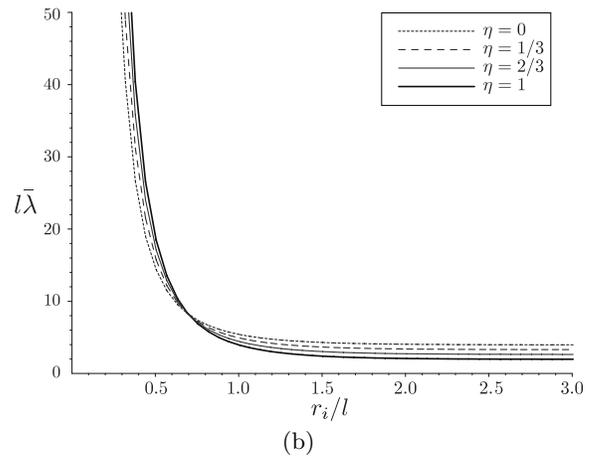} \\
 (b)\\
\end{tabular}
\caption{The tension for a de Sitter brane instanton
 with $\eta=0, 1/3, 2/3$, and $1$.
The figures (a) and (b) correspond to the cases of a plus  and a minus
branches, respectively.
 $d\bar{\lambda} /dr$ is always negative because $2 \pm \eta>0$ and
 $1 \mp \eta>0$ for the range of $0 \leq \eta <1$.
  Since $\bar{\lambda}$ monotonically decreases with respect to $r$
 and approaches the critical tension as $r$ gets large, 
  $\triangle\bar{\lambda}$ is always positive.}
\label{tension-dS}
\end{center}
\end{figure}

In our case, if we take the limit of $r \rightarrow \infty$,
the brane approaches to the 4D Euclidian flat space
because the radius of the brane $(b(r))$
becomes infinitely large
and the curvature of the brane  (S$^4$ manifold)  vanishes.
In this limit, the tension of the  brane (\ref{tens1}) is 
\begin{eqnarray}
 \bar{\lambda}_{i, {\rm cr}}=(-1)^i \frac{2}{l_\pm} (2\pm \eta)
 .  
\label{cri}
\end{eqnarray}
This value is consistent with 
the above generalized Randall-Sundrum tuning
condition (\ref{cr_ten}).
Using this critical tension, 
the tension of a positive-tension brane is divided into two
parts
$\bar{\lambda} = \bar{\lambda}_{\rm cr} +\triangle\bar{\lambda}$. 
It turns out that 
$\triangle\bar{\lambda}$ is always positive
because $\bar{\lambda}$ decreases monotonically 
with respect to $r$  for
$0 \leq \eta <1$ 
(see Fig. 
\ref{tension-dS}).
Hence, this brane has always 
a positive effective cosmological constant, that is, de Sitter brane.

We calculate the action in order to discuss which state 
is most plausible when the brane universe is created. 
The total Euclidian action for this solution is calculated as
\begin{eqnarray}
 S_E &=&   -\frac{V_4^{(+)}}{\kappa_5^2} l_\pm^3
 \left\{
  \left[
 l_\pm^3 (\Lambda l_\pm^2 +9) \sinh 
     \left(\frac{r_2}{l_\pm} \right)
   \cosh 
     \left(\frac{r_2}{l_\pm} \right)
\right.\right.
  \nonumber \\
&&
\left.\left.+3l_\pm^3(\Lambda l_\pm^2+5 )
     \left(\frac{r_2}{l_\pm} \right)
   \right] 
 - \left[r_2 \rightarrow r_1 \right] 
 \right\},
\end{eqnarray}
where we have used the relation
$\Lambda l_\pm^4+6 l_\pm^2-12 \alpha=0$.
Here  $V_4^{(+)}=8\pi^2/3$
because of 
$S^4$ manifold.     
We rewrite the action using the variable $\eta$, 
\begin{eqnarray}
  S_E &=& -\frac{3V_4^{(+)}}{\kappa_5^2} l_\pm^3
 \left\{
  \left[
   (2 \mp \eta) \sinh 
     \left(\frac{r_2}{l_\pm} \right)
   \cosh 
     \left(\frac{r_2}{l_\pm} \right)
 \right.\right.
  \nonumber \\
&&
\left.\left.
 +(2 \mp 3\eta)
     \left(\frac{r_2}{l_\pm} \right)
   \right] 
 - \left[r_2 \rightarrow r_1 \right] 
 \right\}.
\end{eqnarray}
For single brane instanton, the action is given by
replacing $r_2$ and $r_1$ with $r_0$ and $0$, respectively.
 
Those actions do not have any minimum value,
and get small when the distance between two branes
or 
the size of the brane 
becomes large. 
Although we can claim that the brane universe may
be created
as large as possible, we cannot predict the initial size.

The evolution of a brane after creation is given by
analytic continuation of the Euclidean metric
\begin{eqnarray}
 ds_E^2=dr^2
+l^2 \sinh^2 (r/l_\pm)
 \left(d\chi^2 +\sin^2 \chi d\Omega_{(3)}^2 \right)
\end{eqnarray}
by the Wick rotation, 
where $d\Omega_{(3)}^2$ is the metric of the 3-sphere.
It is done by substituting $\chi \rightarrow i H t+ \pi/2 $, 
which leads to
\begin{eqnarray}
ds^2&=&dr^2+(l_\pm H)^2 \sinh(r/l_\pm) \nonumber\\ &&
\times[-dt^2+H^{-2}\cosh^2(Ht)d\Omega_{(3)}^2 ],~~~~~
\end{eqnarray}
where $H \equiv l_\pm \sinh (r_i/l_\pm)$ $(r_i=r_0$ or $r_2)$ is the
radius of a brane.
After the creation of this spacetime, the universe inflates.
If $\triangle\bar{\lambda}$ is given by some potential of a scalar field
and will decrease to zero, 
inflation will end (see \S. 
\ref{trace-anomaly}).

\subsection{flat brane instanton $(k=0)$}
In the case of a flat brane ($k=0$),
the solution of Eq. (\ref{eq1}) is
$ b(r)=b_0 e^{\pm r/l_\pm}$,
where $b_0$ is an integrating constant.
Due to the $Z_2$ symmetry, 
we consider only the plus sign without loss of generality, i.e. 
\begin{eqnarray}
 b(r)=b_0 e^{r/l_\pm}\,. 
\label{sol2}
\end{eqnarray}
We can construct an instanton solution in the same way
as the previous case.
We can impose the junction conditions at the brane boundaries
 ($r=r_1$ and $r_2$, or
$r_0$).
This solution, however, does not satisfy no-boundary boundary condition,
because $b(r)$ dose not vanish at any point $r$.
Thus we cannot construct a single brane instanton solution. 
Here we consider  only a two-brane model.

The tension of $i$-th brane is determined by 
the junction condition (\ref{junc}) as
\begin{eqnarray}
 \frac{b'}{b}\left(3-4\alpha \frac{b'^2}{b^2}\right)
 =\mp
\frac{\bar{\lambda}_i}{2}. 
\label{junc1}
\end{eqnarray}
Substituting Eq. (\ref{sol2}) into Eq. (\ref{junc1}),
\begin{eqnarray}
 \bar{\lambda}_i^{(\pm)} = (-1)^i {2\over {l_\pm}}(2 \pm \eta)\,.
\end{eqnarray}
This tension  is independent of the position
of a brane and is the same as the critical tension
(\ref{cri}).

As for the Euclidean action of this solution,
we find
\begin{eqnarray}
 S_E = -\frac{2V_4^{(0)}b_0^4(2 \mp 3\eta)}{\kappa_5^2~l_\pm}
\left[ e^{4 r_2/l_\pm}- e^{4r_1/l_\pm}
\right].
\end{eqnarray}
Here $V_4^{(0)}$ is the volume of 4D torus.
This action does also not have any minimum value
with respect to the distance between two branes.
Then we cannot predict the initial size of the universe.
Note that we can obtain one-brane RS II model in the limit of
$r_1\rightarrow-\infty$. In that case, the total action
$S_E$ is still finite. Therefore, an instanton solution with one flat brane 
exists.

The evolution of a brane universe after creation is also given by
analytic continuation of the Euclidean metric
\begin{eqnarray}
 ds_E^2=dr^2+b_0^2 e^{2 r/l_\pm}
 \left(d\tau^2 +dx^2+dy^2+dz^2 \right)
\end{eqnarray}
by the Wick rotation. 
Substituting $\tau \rightarrow i t$, 
we obtain
\begin{eqnarray}
 ds^2=dr^2 +b_0^2 e^{2  r/l_\pm} [-dt^2+dx^2+dy^2+dz^2 ]\,.
\end{eqnarray} 
We recover the 4D Minkowski spacetime.

\subsection{anti de Sitter brane instanton $(k=-1)$}
In the case of an `open' brane model ($k=-1$),
the solution of Eq. (\ref{eq1}) is given as
\begin{eqnarray}
 b(r)=l_\pm \cosh \left(\frac{r}{l_\pm} \right)\,.
 \label{sol3}
\end{eqnarray} 
We impose the junction condition at the boundary ($r=r_1$ and $r_2$). 
As the same reason for a flat brane model,
this solution does not provide a single brane model.

The tension of $i$-th brane is determined by the junction condition 
(\ref{junc}) as
\begin{eqnarray}
\frac{b'}{b}\left( 3 -4 \alpha \frac{b'^2}{b^2} -12\alpha \frac{1}{b^2} 
 \right)= \mp \frac{\bar{\lambda}_i}{2}. \label{junc2}
\end{eqnarray}
Substituting Eq. (\ref{sol3}) into Eq. (\ref{junc2}),
we obtain 
\begin{eqnarray}
 \bar{\lambda}_i^{(\pm) }&=& (-1)^i \frac{2}{l_\pm}
 \left[(2 \pm \eta) \frac{\sinh(r_i/l_\pm)}{\cosh(r_i/l_\pm)}
\right.
\nonumber \\
&& 
\left.
~~~~~~~~~~
- 2(1 \mp \eta)\frac{\sinh(r_i/l_\pm)}{\cosh^3(r_i/l_\pm)}
 \right]. 
\label{tens2}
\end{eqnarray}
In the limit of $r \rightarrow \infty$, 
we recover the critical tension  (\ref{cri}),
since the curvature of a brane vanishes.
Furthermore, if we divide the tension (\ref{tens2}) into two parts;
$\bar{\lambda} =\bar{\lambda}_{\rm cr}+\triangle\bar{\lambda}$,
we find that   
$\triangle\bar{\lambda}$  is always negative for
$0 \leq \eta < 1$. 
Since this tension gives a negative effective cosmological constant
on the brane,
the brane is anti-de Sitter spacetime.
\begin{figure}[ht]
\begin{center}
\begin{tabular}{c}
 \includegraphics[height=5.5cm]{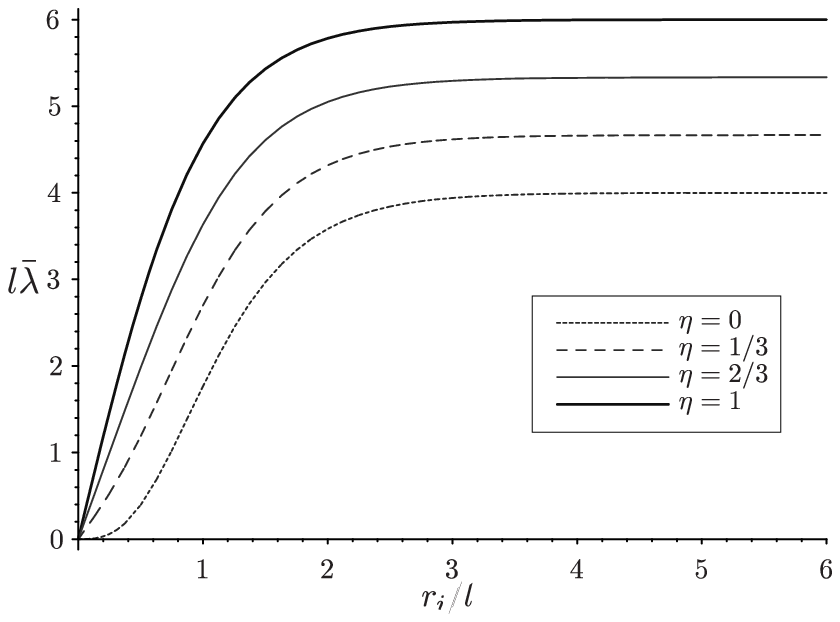} \\ 
 (a)\\[5mm]
 \includegraphics[height=5.5cm]{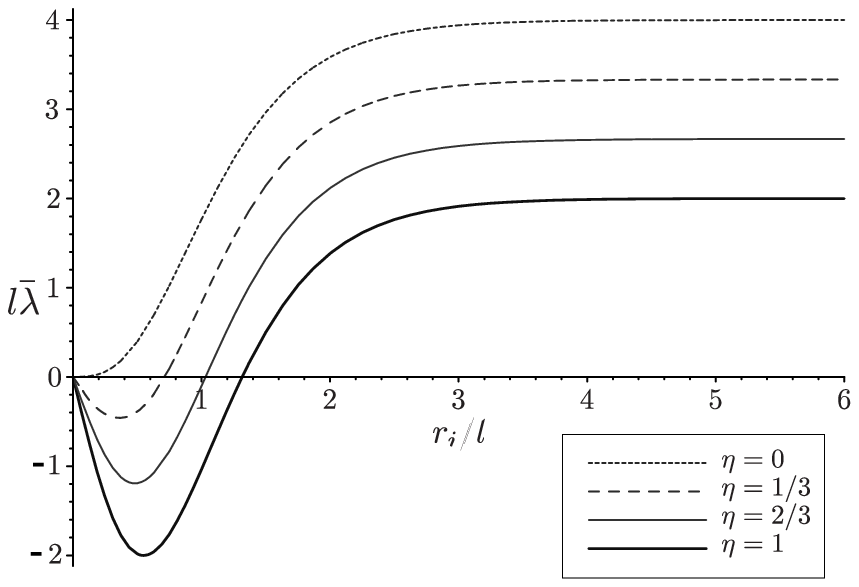} \\
 (b)\\
\end{tabular}
\caption{The tension for an open brane instanton 
 with $\eta=0,1/3,2/3$, and $1$.
Figs. (a) and (b) depict the tension of a plus  and a minus branches, 
respectively.
 $d\bar{\lambda} /dr$ changes sign from negative to positive 
 and $\bar{\lambda}$ approaches the critical tension in the limit of 
$r \rightarrow \infty$. Since $\bar{\lambda}=0$ at $r=0$,
 $\triangle\bar{\lambda}$ is
always negative.}
\label{tension-open}
\end{center}
\end{figure}

The total action ($S_E =S_E^{\rm bulk}+S_E^{\rm brane}$) is
given by
\begin{eqnarray}
 S_E &=&  -\frac{3V_4^{(-)}}{\kappa_5^2} l_\pm^3
 \left\{ \left[
 (-2 \pm \eta) \cosh \left(\frac{r_2}{l_\pm}\right)
 \sinh  \left(\frac{r_2}{l_\pm}\right) 
 \right.\right.
\nonumber \\
&&\left.\left.
~~~~~~~~+(2 \mp 3\eta) \left(\frac{r_2}{l_\pm}\right) \right]
 - [r_2 \rightarrow r_1]
 \right\}
\end{eqnarray}
Here $V_4^{(-)}$ is the volume of 4D manifold with $k=-1$.
Again we do not have any minimum in this action.
We cannot give any prediction for a created brane spacetime.

To discuss the evolution of the brane universe after creation,
we have to perform
analytically continuation of the Euclidian space;
\begin{eqnarray}
 ds_E^2=dr^2+l_\pm^2 \cosh^2 (r/l_{\pm}) ds_{E,4}^2
\end{eqnarray}
with 
\begin{eqnarray}
ds_{E,4}^2=d\chi^2+\sinh^2
 \chi\left(d\psi^2 +\sin^2\psi d\Omega_{(2)}^2\right) \,.
\end{eqnarray}
By the double Wick rotations, i.e. 
$\chi \rightarrow i(t+\pi/2)$ and
$\psi \rightarrow i\phi$ \cite{Witten82},
we obtain the metric of the brane as
\begin{eqnarray}
 ds^2_4=-dt^2+\cos^2 t
 \left(d\phi^2+\sinh^2 \phi d\Omega_{(2)}^2 \right)\,,
\end{eqnarray}
which gives the AdS spacetime.

\section{Trace Anomaly}
 \label{trace-anomaly}
In the previous section, although
we can find instanton solutions for several situations,
we cannot provide the most plausible state of the brane universe
because the action does not give any minimum value.
If we choose the critical tension to obtain zero cosmological constant,
only a flat brane instanton is possible.
The size of the brane universe, however, is not determined.
Any distance between two brane is possible.
We again loose the predictability. 
In this section we discuss another effect which may 
give a prediction of the initial state of the brane universe.
In a curved space time, we know  that even in the absence of 
classical gravitational action, quantum fluctuation of 
 matter fields provides a nontrivial gravitational action
 through a trace anomaly term
$\langle \tau_{\mu \nu} \rangle$. 
In the case of free, massless, and conformally invariant fields,
these quantum corrections take a simple form \cite{Birrell-Davies}.
Those term were discussed in the context of inflationary
scenario \cite{Starobinsky80} and 
creation of the universe \cite{Vilenkin85, Hawking-Hetrog-Reall01}.
In \cite{Hawking-Hetrog-Reall01}, assuming the critical
tension, the size of the brane
universe is fixed.
Then, in the present case, we also take into account of the 
contribution of the trace anomaly,
our junction condition is modified as
\begin{eqnarray}
 [B_{\mu \nu} ]^{\pm}
& \equiv&\left[-\frac{b'}{b}\left\{3-4\alpha \left(\frac{b'^2}{b^2} 
  -3 \frac{k}{b^2}\right) \right\}\right]^{\pm} h_{\mu \nu}
\nonumber \\
&=& -\kappa_5^2\left(\tau_{\mu \nu} + \langle \tau_{\mu \nu} \rangle
\right)
\,,
\end{eqnarray}
where  $\tau_{\mu \nu}$ is the energy-momentum tensor of the brane matter fields
 and $\langle \tau_{\mu \nu}\rangle$ is
its trace anomaly term including the tension of the brane, which is given by
\begin{eqnarray}
\langle \tau_{\mu \nu}
\rangle = - \lambda h_{\mu \nu} + {H}^{(1)}_{\mu \nu} +{H}^{(3)}_{\mu \nu}
.
\end{eqnarray}
${H}^{(1)}_{\mu \nu}$ and ${ H}^{(3)}_{\mu \nu}$
take the following forms;
\begin{eqnarray}
 {H}^{(1)}_{\mu \nu}&=& -k_1 \left(
 2R R_{\mu \nu} - \frac{1}{2} h_{\mu \nu} R^2 
 -2 \nabla_{\mu} \nabla_{\nu} R 
 \right.
\nonumber \\
&&\left.
+2h_{\mu \nu}\nabla^{\alpha}\nabla_{\alpha}R \right),
\end{eqnarray}
\begin{eqnarray}
 {H}^{(3)}_{\mu \nu} &=& k_3 \left( 
 -R_{\mu}^{\ \sigma}R_{\nu \sigma} + \frac{2}{3} RR_{\mu \nu}
 +\frac{1}{2}h_{\mu \nu}R^{\sigma \tau}R_{\sigma \tau} 
 \right.
\nonumber \\
&&\left.
   -\frac{1}{4}h_{\mu \nu} R^2 \right),
\end{eqnarray}
where $R$ and $R_{\mu \nu}$ are the 4D scalar curvature
and Ricci tensor respectively. 
The coefficient $k_1$ may not appear in ${\cal N}=4$ super conformal Yang-Mills
theory but can be included  to obtain a successful inflationary scenario.
While $k_3$ is uniquely determined: 
\begin{eqnarray}
 k_3=\frac{1}{2880 \pi^2}(2N_0 +11N_{1/2}+62N_1),
\end{eqnarray}
where $N_0$, $N_{1/2}$, and $N_1$ are the number of 
quantum fields with spins $0$, $1/2$, and $1$, respectively.

We shall include the trace anomaly terms for our instanton solutions.
By using the metric (\ref{Emet}),
\begin{eqnarray}
 R = \pm \frac{12}{b^2}, \quad 
 R_{\mu \nu} = \pm \frac{3}{b^2}h_{\mu \nu}, \quad
 \nabla_{\mu}R=0,
\end{eqnarray}
where the upper (or lower) sign corresponds to the $k=1$  brane instanton (or
$k=-1$ brane one). For a flat brane, the scalar curvature and  Ricci tensor 
vanish.

The trace anomaly terms are then given by
\begin{eqnarray}
 {H}^{(1)}_{\mu \nu} =0,
\end{eqnarray}
\begin{eqnarray}
 {H}^{(3)}_{\mu \nu}  = - \frac{3k_3 }{b^4} h_{\mu \nu}.
\end{eqnarray}
We obtain 
\begin{eqnarray}
 \langle \tau_{\mu \nu} \rangle = - \frac{3k_3}{b^4}h_{\mu \nu}. 
\label{trace1}
\end{eqnarray}
Hence the junction condition is 
\begin{eqnarray}
 \left[\frac{b'}{b}\left\{3-4\alpha \left(\frac{b'^2}{b^2} 
  -\frac{3 k}{b^2}\right) \right\}\right]^{\pm}
 = -\left(\bar{\lambda} + \frac{3\bar{k}_3}{b^4}
\right)
\,,
\end{eqnarray}
where
$\bar{\lambda}=\kappa_5^2 \lambda$ and $\bar{k}_3=\kappa_5^2 k_3$.

\begin{figure}[t]
\begin{center}
\begin{tabular}{c}
 \includegraphics[height=5.5cm]{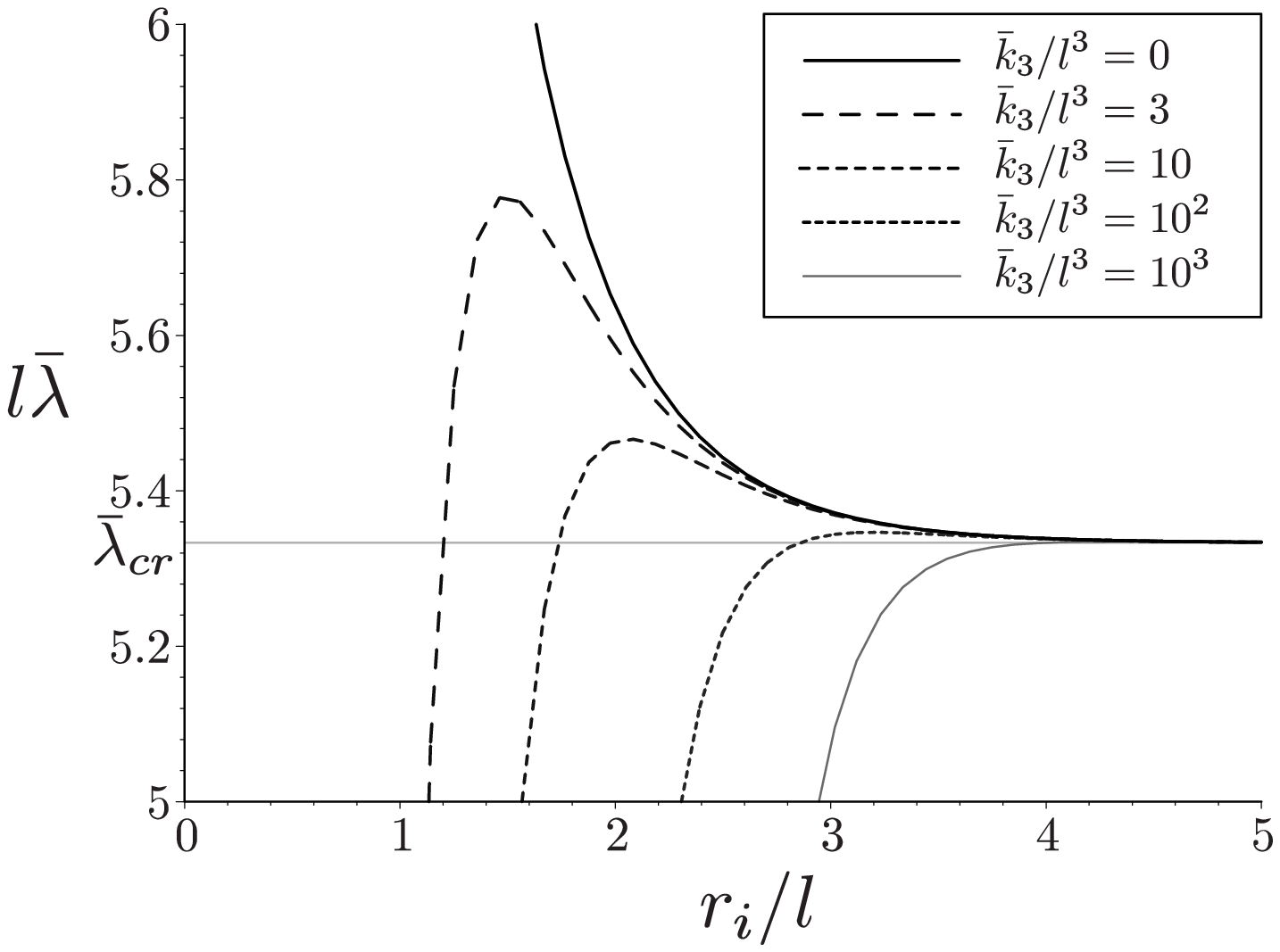} \\ 
 (a) \\[5mm]
 \includegraphics[height=5.5cm]{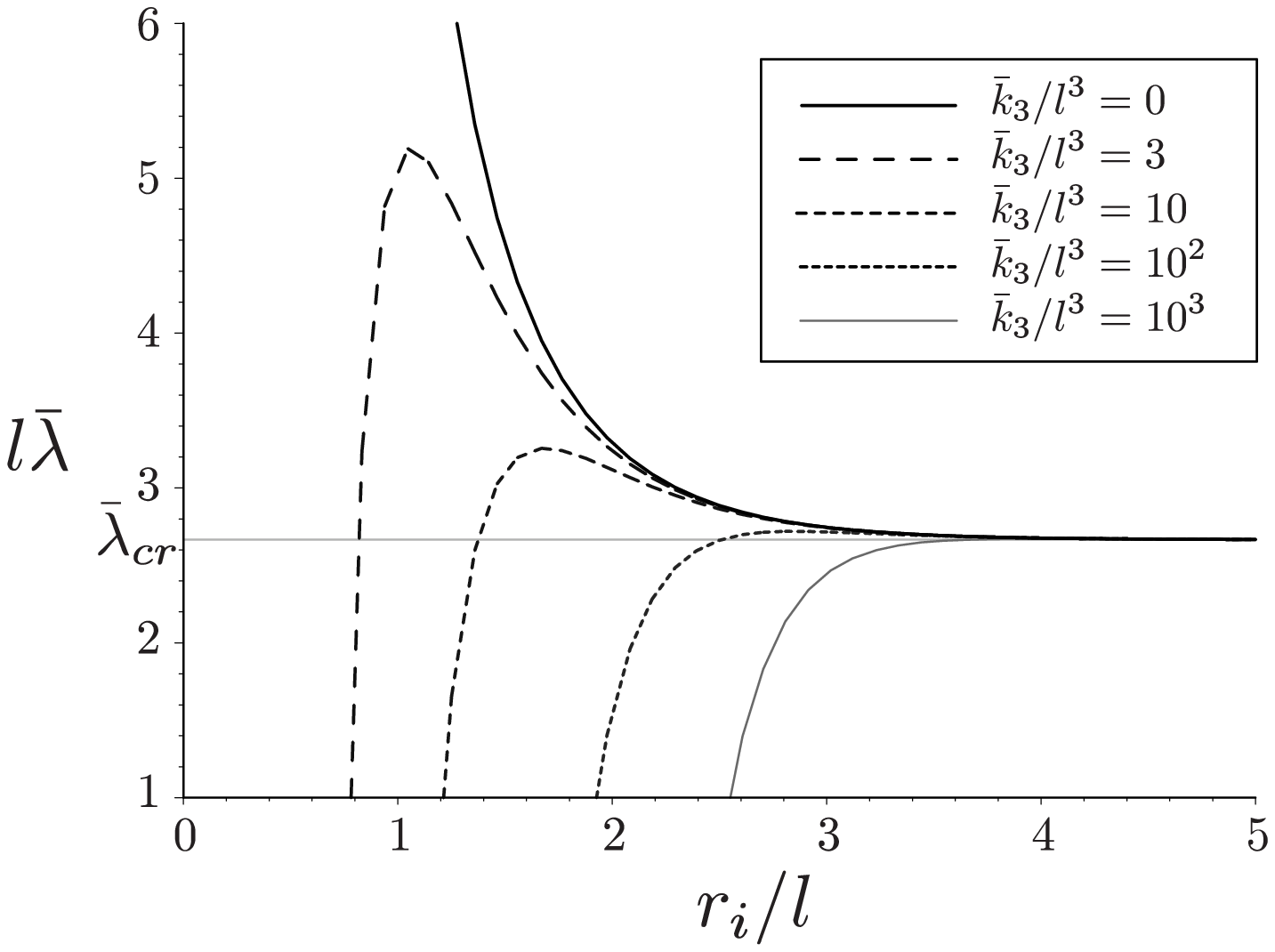} \\ 
 (b)\\
\end{tabular}
\caption{The tension for a de Sitter brane instanton
 with  $\eta=2/3$ and $\bar{k}_3/l_\pm=3, 10, 10^2, 10^3$, and 0,
where $\bar{k}_3=0$ gives the case without the trace anomaly term.   
The figures (a) and (b) correspond to the cases of a plus-  and a minus-
branches, respectively.
Since $\lim_{r_i \rightarrow \infty}[\bar{\lambda}-\bar{\lambda}_{\rm cr}]=0+$,
we find a unique solution at finite radius 
if the tension is the critical value $\bar{\lambda}_{\rm cr}$.}
\label{tension-dS_TA}
\end{center}
\end{figure}

For a positive tension brane (either a single brane or the second brane
of two-brane model),
we obtain 
\begin{eqnarray}
 \bar{\lambda} = 2\frac{b'}{b}\left\{3 -4\alpha \left(\frac{b'^2}{b^2} 
 - \frac{3 k}{b^2} \right)\right\}-  \frac{3\bar{k}_3}{b^4} 
\label{junc-ta}
\,.
\end{eqnarray}
Since a trace anomaly (\ref{trace1}) is always negative 
(or zero for a flat brane),
the tension is always below the value obtained previously.
It turns out that only de Sitter brane with $k=1$
is possible if we adopt the critical tension.
We show the tension in terms of including trace anomaly term in Fig.
\ref{tension-dS_TA}.
\begin{eqnarray}
 \bar{\lambda}^{(\pm)}&=&\frac{2}{l_{\pm}} \left\{(2\pm \eta)+
 \frac{2(1 \mp \eta)}{\sinh^2 (r_i/l_{\pm})} \right\} 
\nonumber  \\
&& -\frac{3\bar{k}_3}{l_{\pm}^4\sinh^4 (r_i/l_{\pm})}\,.
\end{eqnarray}
If the tension is  the critical value (\ref{cri}), we find a unique solution 
with a
finite radius. 

\begin{figure}[ht]
\begin{center}
\begin{tabular}{c}
 \includegraphics[height=5.5cm]{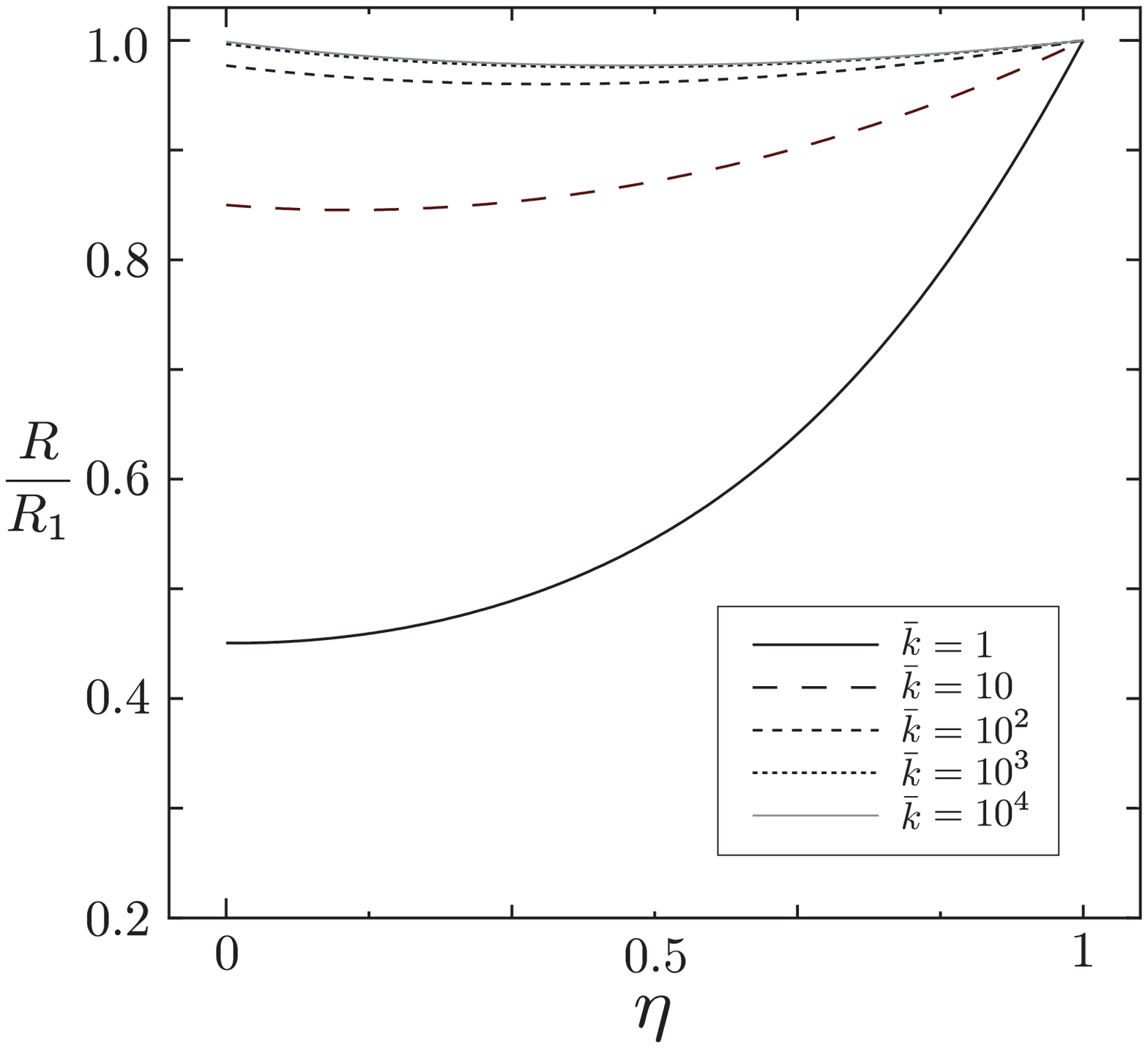} \\
 (a)\\[5mm]
 \includegraphics[height=5.5cm]{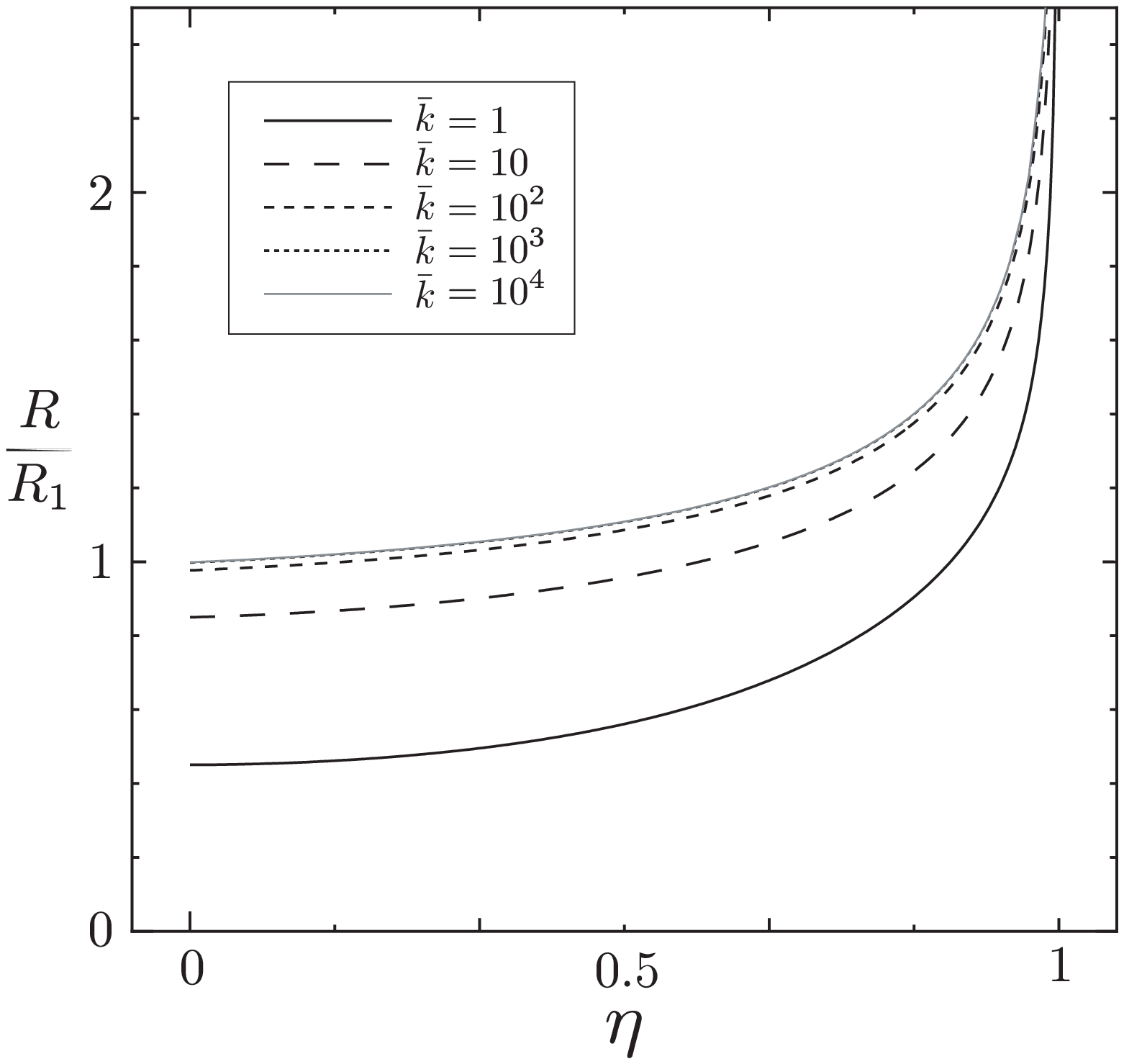} \\
 (b)\\
\end{tabular}
\caption{The radius of a de Sitter brane instanton
with respect to $\eta$.
The figure shows the ratio of the nucleation radius to that without
the Gauss-Bonnet term.
The figures (a) and (b) correspond to the cases of a plus-  and a minus-
branches, respectively.
Here $R$ is the radius of the positive brane : 
$R \equiv l_\pm \sinh(r_i/l_\pm)$ 
and $R_1$ is the radius of the positive brane with $\eta=1$ :
$R_1 \equiv l^{\pm}_1 \sinh (r_i/l^\pm_1)$, 
where $l^\pm_1\equiv l_\pm(\eta=1)$.}
\label{radius_deSitter_instanton}
\end{center}
\end{figure}

For a two-brane model, if the tension on one brane is critical, the other
is not the case. For example, if the tension of the positive tension brane
 ($r_2$) is
critical, a effective cosmological constant on the negative tension brane ($r_1$)
is positive. Conversely, if 
 the tension of the negative tension brane ($r_1$)
is critical, we find AdS universe on the positive-tension brane. 
For a single brane model, 
the radius of the created brane universe is fixed.
 In Fig. \ref{radius_deSitter_instanton}, we depict the ratio of the size of a
single brane with respect to $\eta$ to that without the Gauss-Bonnet term, which
is  obtained by \cite{Hawking-Hetrog-Reall01}.
We find that the difference is not so large.
For the plus branch, the radius of the created universe is 
always smaller than that without Gauss-Bonnet term.
While, for negative branch, it highly depends on the value of $\eta$.
If we take the limit of $\eta \rightarrow 1 (\alpha \rightarrow 0)$, the
radius gets much larger.

Note that since the trace anomaly always vanishes for a flat brane,
the flat two-brane model at any distance is possible.

\section{CONCLUSION} \label{conclusion}
We have presented instanton solutions in the model with a Gauss-Bonnet term.
If a brane is a closed universe,
we find both two-brane and single-brane de Sitter brane instantons.
For a flat brane universe, we can also construct a two-brane instanton
solution.
As for a single-brane model, 
a  compact bulk spacetime with a single brane is not
possible  because a flatness of
the brane is not consistent with a 
no-boundary boundary condition. 
However, the RS II type instanton with a  non-compact bulk spacetime
 is allowed because its Euclidean action is finite.
For an anti-de Sitter brane instanton with negative 
curvature, we can also construct only a two-brane
instanton solution. 
In this case, the RS II type instanton does not exist because the
Euclidean action diverges.

Although we find  instanton solutions in the model 
with Gauss-Bonnet term, we cannot predict the
initial state of the brane universe.
This is because the Euclidian action has no minimum value.
For a flat and an anti-de Sitter brane,
 the volume of a created brane universe
$V^{(k)}_4$ is  not fixed as well.
$V^{(k)}_4$ can be arbitrary.

As for the tension of a brane, 
although the critical tension requires a fine-tuning.
we need such a choice to explain the present small value
of the universe.
Such a tuning could be obtained in some super symmetric  theories
such as Horava-Witten model.
Here we assume such a tuned value, i.e. the critical
tension.
Then we also include 
 trace anomaly 
terms on the brane.
In this case, we can predict the size of the universe.
If we have a single-brane universe with no-boundary boundary
condition,
the created universe is in de Sitter phase and naturally 
evolves into inflationary stage.

According to the recent analysis by Charmousis and
Dufaux \cite{Charmousis-Dufaux03}, 
 the models with a negative-tension brane
show its instability in the Einstein-Gauss-Bonnet theory.
However such a   problem is absent in the non-compact RS II model.
For de Sitter brane instanton with two branes,
this instability may be
smeared out by the existence of 
trace anomaly terms similarly to the inclusion of
induced gravity term on the branes.
Furthermore in order to predict the initial state of created brane universe,
we need to include other important effects such as the Casimir energy, which are
not taken into account here. These issues are left for a future study.

\acknowledgments

We would like to thank S. Mizuno, N. Okuyama  and T. Torii 
for useful discussions and comments.
This work was partially supported by the Grant-in-Aid for
Scientific Research  Fund of the MEXT
 (No. 14540281) and by the Waseda University Grant for
Special Research Projects and  for The 21st Century 
COE Program (Holistic Research and Education Center for Physics 
Self-organization Systems) at Waseda University.

\appendix

\section{Euclidean Action}
Here we present the Euclidean action.
First, we calculate a scalar curvature and a Gauss-Bonnet term 
using the metric (\ref{Emet}) as
\begin{eqnarray}
 R &=& 12\left(\frac{k-b'^2}{b^2}\right)-8\frac{b''}{b},   \\  
 {\mathcal L}_{GB} &=& 24 \left(\frac{k-b'^2}{b^2}\right)^2-96
\left(\frac{k-b'^2}{b^2}\right) {b''\over b}.
\end{eqnarray}
By using Einstein equations  (\ref{eqm1}), we rewrite the bulk action 
(\ref{eqm2}) as 
\begin{eqnarray}
 S_E^{\rm bulk} &=& - \frac{4}{\kappa^2_5} \int_{\mathcal M} dx^5 \sqrt{g}
  \left[ \Lambda+ 2\frac{b''}{b} -3\frac{k-b'^2}{b^2} \right] 
\nonumber
\\
  &=& -  \frac{8V_4^{(k)}}{\kappa^2_5} \int_{r_1}^{r_2} dr
  \left[
\Lambda
+2\frac{b''}{b} -3\frac{k-b'^2}{b^2}
\right].
\label{action_bulk}
\end{eqnarray}  
In a single-brane model,
we just replace 
the integration range  $[r_1,r_2]$   with $[0, r_0]$.
In the second line of Eq. (\ref{action_bulk}), we put the factor 2
because an instanton is constructed from
 two copies of a patched manifold with
region
$[r_1, r_2]$ (or $[0,r_0]$).

We then calculate the  surface term, i.e.
\begin{eqnarray}
&& K+ 2\alpha \left(J - 2G^{\rho \sigma}K_{\rho \sigma} \right) 
\nonumber \\
 &&~~~= 4\left(\frac{b'}{b}\right)\left[1 - 4\alpha \left(\frac{b'^2}{b^2} 
-\frac{3k}{b^2}  \right)\right]\,,
\end{eqnarray}
where we have used $J_{\mu \nu}= -2 \left(b'/b\right)^3 h_{\mu \nu} $ and $  
J =  -8
\left(b'/b\right)^3$.
Using above results, we rewrite the boundary action
\begin{eqnarray}
 &&
S_E^{\rm brane}=\frac{1}{\kappa_5^2} \int_{\partial {\mathcal M}_i}
d^4x 
\sqrt{h}
 \nonumber \\
&&~~\times
 \left\{ 4\left[\left(\frac{b'}{b}\right)\left\{1 - 4\alpha
\left(\frac{b'^2}{b^2}  -\frac{3k}{b^2}  \right)\right\}\right]^{\pm} 
+\bar{\lambda} \right\}\,.
\label{S^b2}
\end{eqnarray}
Using the junction condition (\ref{junc}), (\ref{S^b2}) gives
\begin{eqnarray}
 S_E^{\rm brane}&=& \frac{1}{\kappa_5^2} \int_{\partial {\mathcal M}_i}
d^4x 
\sqrt{h}
\nonumber \\
&\times&
 \left[ \left(\frac{b'}{b}\right)\left\{1 - 12\alpha \left(\frac{b'^2}{b^2} 
-\frac{3k}{b^2}  \right)\right\}\right]^{\pm}  .
\end{eqnarray}
Imposing $Z_2$ symmetry ($b'_{-}=-b'_{+}$), 
the Euclidean action for a two-brane model is rewritten as 
\begin{eqnarray} 
 S_E^{\rm brane}&=& \frac{2V_4^{(k)}}{\kappa_5^2} 
\left\{   \left(\frac{b'_{+}}{b}\right)\left[ 1 -12\alpha\left(
\frac{b'^2_{+}}{b^2}
   - \frac{3k}{b^2}\right)  \right]\right\}_{r=r_1}
 \nonumber \\
&
 -&\frac{2V_4^{(k)}}{\kappa_5^2} \left\{
  \left(\frac{b'_{-}}{b}\right)\left[ 1 -12\alpha\left(
\frac{b'^2_{-}}{b^2}
   -  \frac{3k}{b^2}  \right)\right]\right\}_{r=r_2}.
\nonumber 
\\~
\end{eqnarray}
For a single-brane model, 
\begin{eqnarray}
 S_E^{\rm brane}&&=  -\frac{2V_4^{(k)}}{\kappa_5^2} 
\left\{
 \left(\frac{b'_{-}}{b}\right)\left[ 1-12\alpha\left(
\frac{b'^2_{-}}{b^2}
   -  \frac{3k}{b^2}  \right)\right]\right\}_{r=r_0}.
\nonumber 
\\&&~
\end{eqnarray}
Finally we obtain the total action by $S_E^{\rm bulk}+S_E^{\rm brane}$.


\end{document}